\documentclass[spanish,english]{article}
\usepackage[T1]{fontenc}
\usepackage[latin9]{inputenc}
\usepackage{geometry}
\usepackage{amsmath}
\usepackage{amssymb}
\usepackage{cancel}
\usepackage{stmaryrd}
\usepackage{esint}
\usepackage{babel}
\addto\shorthandsspanish{\spanishdeactivate{~<>}}

\begin{document}
\title{Spacetime Relationalism in GR and QG}
\author{{\normalsize{}Alejandro Ascárate}\thanks{FaMAF-UNC (Córdoba, Argentina). Email: aleazk@gmail.com}}
\maketitle
\begin{abstract}
We provide here a philosophical basis for \cite{key-4,key-5}, based
on the notion of spacetime relationalism. We argue that the view which
is more cleanly compatible with GR is that in which the spacetime
manifold is a relational (or ``emergent'') property of the aggregate
or physical sum of gravitational field portions, field that is considered
material and for which spacetime geometry (in the sense of the metric
properties) is one of its material properties. All of the previous
philosophical notions are defined in relation to a clear ontology
of material individuals, briefly described in the appendix. Furthermore,
this fits well with the phase space of GR, thing that also may inform
how to proceed in the quantum realm.
\end{abstract}
$\;$

\subsection*{{\normalsize{}1.1 Spacetime Relationalism in GR}}

$\;$

Regarding space and time, Newton had a peculiar position. In his words:
``absolute space, in its own nature, without regard to anything external,
remains always similar and immovable. Relative space is some movable
dimension or measure of the absolute spaces; which our senses determine
by its position to bodies: and which is vulgarly taken for immovable
space... Absolute motion is the translation of a body from one absolute
place into another: and relative motion, the translation from one
relative place into another... Absolute, true and mathematical time,
of itself, and from its own nature flows equably without regard to
anything external, and by another name is called duration: relative,
apparent and common time, is some sensible and external (whether accurate
or unequable) measure of duration by the means of motion, which is
commonly used instead of true time ...''- Isaac Newton. That is,
if all matter in the universe suddenly disappears, space and time
still would exist, according to this view. In this way, we would need
to introduce a second substance in our ontology, a ``Spatio-Temporal
Substance'', and which is, of course, different and independent from
standard material substance. This point of view is called ``Space-time
substantivalism''. Note, however, that he gives a role to this substance:
it defines absolute inertial motion.

$\;$

Leibniz, among others, had quite a different view, but which is more
appropriate for the notion of Reality adopted in the Appendix. Regarding
space, it states that Things are presented in reality not in an arbitrary
way, but by keeping \emph{separation} relations\footnote{Proto-physically speaking, we can think of its physical meaning as
resulting from the axiom that any signal from one Thing takes a \emph{finite}
time duration (see next footnote) to reach a second one (not superimposed
and separated from the first.) \emph{Distance} values, in this scheme,
are related to the actual different quantitative values of these time
durations.} and, then, we could say that, given more than one thing, then a property
of this system emerges and which is just physical space. Regarding
time, note that, given a Thing, it can present, at least in principle,
different states. Indeed, we will postulate that all Things are \emph{Mutable};
that is, they \emph{have} at least two \emph{different possible} \emph{physical
states} (e.g., states $(q,p)=(2,5)$ and $(q,p)=(1,2)$ in classical
mechanics.) Then, we postulate that there is at least one thing in
the universe that \emph{Changes}: that is, for which its state effectively
mutates from one to another. The new point of view is simply that
time is (actually, a \emph{layered} notion, consisting in) Mutability
and Change (note that only Mutability is analogous to Separation)\footnote{Proto-physically, time \emph{duration} is related to a \emph{strict
ordering} in the successive events that comprise this change, i.e.,
it's yet another new layer in the concept of time (as distance is
in space.) That is, the set of all points of space is a \emph{different
thing} and \emph{prior} to the metric one \emph{puts into it}. For
the set of points itself, we take the emergent/relational view. In
this sense, one thing is \emph{mutability} and \emph{change}, and
other very different is the \emph{duration} of this change; analogously,
one thing is \emph{separation} and other is the actual \emph{distance}
that separates two points. In both cases, mutability-change and separation
are \emph{previous} to, respectively, duration and distance.}. The advantage of this point of view is that, by postulating the
existence of mutability, change, and separation (and, in particular,
as phenomena tied intrinsically to Things) in the way we did, we then
only need the set of Things $\Theta$ in our ontology, space and time
being now emergent properties from Reality $R$ (in $\Theta$), i.e.,
no new substance is needed. The key difference is, of course, that
if all the Things in the universe suddenly disappear, then so do space
and time. This point of view is called ``Space-time relationalism''.
In Leibniz's words: ``I have said more than once that I hold space
to be something merely relative, as time is; that I hold it to be
an order of coexistents, as time is an order of successions'' (note,
however, that Newton also mentions this notion, but he says it ``is
vulgarly taken for immovable space''.\footnote{Newton and Leibniz also had different views regarding locality. Newton
considered valid the principle of locality or individuation by separation,
while Leibniz didn't (since he believed that two indistinguishable
things must actually be the same; this even if they are separated,
because, being a relationalist regarding the nature of space, he doesn't
consider the different positions of the particles as a property that
distinguishes them, since position would not be an intrinsic and individual
property of each particle, like, say, spin, but an emergent property
of the system of two particles.) Not that there seem to be some similarities
with the non-locality established by quantum entanglement.}) 

$\;$

Now, we adopt as a criterion for the existence of a Physical Material
Field in spacetime if there are accelerations on particles that \emph{cannot}
be \emph{globally} eliminated by passing to another reference frame
(although, it may be \emph{locally}), since, if they could, in that
case it would just be an innocuous artificial acceleration caused
by the acceleration of the reference frame itself and nothing more
than that. Electromagnetic fields obviously satisfy this criterion
and they also carry energy; thus, they are material entities. But
also gravitational accelerations satisfy this criterion: they can
be locally eliminated but never globally, this is just Einstein's
equivalence principle. Thus, they have to be caused by a Physical
Material Field, which we call the Gravitational Field. In each spacetime
region (actually, infinitesimal) in which they are eliminated, the
spacetime geometry is flat. But, since the elimination cannot be done
globally, this means that the global metric cannot be flat. In this
way, we can characterize a gravitational field by a curved spacetime
geometry. We could also say that \emph{spacetime geometry is actually
a property of the gravitational field}, since it depends on the configuration
of the latter\footnote{That is to say, the geometry that was always studied, both from the
physical and mathematical point of view, was always really the Gravitational
Field itself. When Riemann mathematically studied geometry, we could
say that, in retrospect, what he was studying was simply the mathematical
structure of the Gravitational Field. And this because mathematical
geometry always had its basis in physical geometry, instead of being
a mere mathematical abstraction. In this way, and clarifying the fact
that behind it is the Gravitational Field, we will call geometry to
all properties of this field having to do with issues such as areas,
volumes, durations, and related concepts, such as the Levi-Civita
covariant derivative, the curvature, etc.}. This is the actual content of the theory: spacetime geometry becomes
a property\footnote{In the Ashtekar variables $\left[A_{a}^{i},E_{i}^{a}\right]$ formulation
of General Relativity, one can see this in an explicit way since the
area of a surface or region $S$ is given by $a_{S}\left[E_{i}^{a}\right]=\int_{S}\sqrt{\sum_{j=1}^{3}E_{j}^{a}E_{j}^{b}m_{a}m_{b}}\,\boldsymbol{e}_{S}$,
which is, clearly, a functional $a_{S}:M\,\longrightarrow\mathbb{R}$
on the phase space $M$ of the field described by it to the real numbers,
that is, an authentic property of the field.} of a material\footnote{The field equations predict gravitational waves, i.e., perturbations
in the gravitational field, and, therefore, in spacetime geometry,
that travel through space carrying energy extracted from the system
which emitted them. Thus, according to this theory, the gravitational
field indeed has energy and therefore it's matter.} entity. Once he introduced this setup, Einstein realized that the
motion of a massive body affected by the gravitational field alone
is such that its spacetime trajectory is a geodesic of the curved
spacetime geometry (i.e., taking advantage of the mentioned identification,
one can describe gravitational notions in terms of metric variables.)
Indeed, the geodesic equation equates the coordinate acceleration
to a geometry-dependent term (and, of course, that doesn't depend
on the mass of the body) that, if the geometry is curved, cannot be
transformed away globally. If the geometry is flat, it can be transformed
away globally, and, in this way, the non-accelerated spacetime trajectory
of a free particle in flat spacetime is also a geodesic. Thus, absolute
inertial motion is related to absolute geodesity, i.e., it depends
on the geometry of spacetime; but, since, in General Relativity, geometry
is a property of the Gravitational Field, then, in a rather curious
twist, Newton's absolute substantial space, the one that defined inertial
motion, was just the Gravitational Field! \cite{key-2} (note that
a geodesic can be seen as a spacetime trajectory of constant and non-zero
$3-$velocity in one, perhaps local, inertial frame and zero in another,
that is, inertial motion is not defined here in relation to movement
with respect to some absolute space in absolute rest, as Newton conceived
it and which is its weak point, but in terms of geodesity, which is
unaffected by a change of frames, that is, it's compatible with the
absolute relativity of velocity; what connects this with Newton is
that he perceived that a new substance was needed to define inertial
motion and he just postulated it in the manner and the means which
were accessible to him at his time, but which later resulted to be
the Gravitational Field in the more precise, both mathematically-conceptually
as well as philosophically, description of General Relativity, which
was informed by later discoveries in experimental-theoretical physics
as well as pure mathematics.) Furthermore, there's actually a phenomenom
in General Relativity, called frame-dragging, in which the geodesics
are dragged along the rotation of a very massive object; in this way,
the notion of inertial frame is also being dragged, hence the name,
and this is possible because this notion is really dependent on the
state of the Gravitational Field around the massive object (which
affects this field.) This effect has been experimentally observed
through gyroscopes in orbit around Earth\footnote{Note that there's ontological continuity between Newton's theory and
General Relativity via this reinterpretation of the former's absolute
space. This type of continuities support the notion that there's a
Reality composed of elements and properties and which we can discover
in more detail as our investigation of it advances.}. 

$\;$

In this way, by absorbing Newton's absolute spacetime into the Gravitational
Field, General Relativity seems to suggest or call for a relationist
theory of spacetime\footnote{Although, of course, the substantivalist could still claim that there
seems to be some sort of ``asymmetry'' regarding the respectives
roles of the Gravitational Field and the other matter fields, the
former being related to the metric geometry of spacetime, and even
serving as some sort of ``background'' (more on this below) due
to its special appearence in the equations of the other fields via
the Levi-Civita connection of the derivative, while the other fields
do not seem to be related to geometry in this way. But the relationist
could counter that the Gravitational Field carries energy, and so
it's definitely a type of standard matter substance, rather than a
new kind of substance, and that the Levi-Civita connection is just
a part of the very complex way in which this field interacts with
the other matter fields (the other part being, of course, the Einstein
field equations.)}. Thus, the ontological picture one gets in General Relativity is
the following: the elements of $\Theta_{GR}$ are \emph{only fields}
(gravitational and non-gravitational), from the Reality given by $\Theta_{GR}$
emerges spacetime as a relational structure, and, furthermore, this
structure carries a metric which is actually a property of one of
these fields (the Gravitational Field, of course.) 

$\;$

To make a precise theory that explains how space and time emerge from
Things, and what their physical meanings are, is a task for proto-physics.
And, indeed, the relational sketch we made here can actually be made
more precise \cite{key-1}. We only consider all the \emph{atomic}
or \emph{minimal} Things, that is, those Things which cannot be further
decomposed as the aggregation of other Things. First, we assume Mutability,
that is, that the state\footnote{Recall that any state $s$, in this framework, is such that \emph{all}
properties of the Thing have defined and fixed (for that state) real
values.} space of a Thing has more than just one state (in this way, a process
is a change from one state to another, and, thus, a change in the
values of the properties of the Thing.) Before continuing, note that,
since the actual Universe is just a single one and fixed, the set
of all Things is also one and fixed, since it represents that Universe;
furthermore, the \emph{proper history}\footnote{Which we take as unique and whose points consists in the Thing in
all its allowed legal states once fixed an initial one.} of each of these Things is not a generic, variable one, it's also
one and fixed once and for all. Now, if we assume as ontological hypothesis
that the states of a Thing along its proper history $h$ have a \emph{partial
order} $\leq$ and that this order is \emph{continuous} (that is,
if $h$ is divided into two subsets $h_{p}$ and $h_{f}$ such that
every state in $h_{p}$ temporally precedes any state in $h_{f}$,
then there exists only one state $s_{0}$ such that $s_{1}\leq s_{0}\leq s_{2}$,
where $s_{1}\in h_{p}$ and $s_{2}\in h_{f}$), then it's known from
standard analysis that there's a \emph{bijection} $T:h\,\longrightarrow\mathbb{R}$
such that $h_{1}=\left\{ s(\tau)\diagup\tau\in\mathbb{R}\right\} $,
$s_{0}=s(0)$, and $s_{1}=s(1)$, where the \emph{unit change }$(s_{0},s_{1})$
is arbitrary. Thus, we get \emph{a metric on} $h$ and this metric
is the \emph{proper duration} or \emph{proper time}. What we have
done here is to derive (and, in this way, give \emph{physical meaning})
the notion of metric duration \emph{from ontological axioms about
Things and properties of their set of states}. Now, assume that \emph{any}
pair of Things \emph{interact}. Suppose Thing $x$, whose history
is $h(\tau_{x})$, interacts with Thing $y$, whose history is $h(\tau_{y})$,
that this action starts from $x$ at $\tau_{x}^{0}$, then reaches
$y$ and disturbs it, and then reflects back to $x$, which gets disturbed
at $\tau_{x}^{1}$ when the \emph{reflex action} reaches it. We postulate
that for any two \emph{separated} things, there \emph{always exists
a minimun positive bound for the interval} $(\tau_{x}^{1}-\tau_{x}^{0})$
defined by the reflex action after considering \emph{all} the possible
types of interactions. We define that $\tau_{y}^{0}$ is\emph{ simultaneous}
with $\tau_{x}^{1/2}=\frac{1}{2}(\tau_{x}^{1}+\tau_{x}^{0})$. Of
course, this means that Things $y$ and $x$ in their states at $\tau_{y}^{0}$
and $\tau_{x}^{1/2}$, respectively, are Separated. We now define
a \emph{distance} $d$ between $\tau_{y}^{0}$ and $\tau_{x}^{1/2}$
by $d(x,y)=\frac{c}{2}\mid\tau_{x}^{1}-\tau_{x}^{0}\mid$, where $c$
is some universal constant (we actually postulate that $d$ is a distance
on the equivalence class $E$ of states simultaneous to $\tau_{y}^{0}$.)
Geometric space $E_{G}$ is the completion of the metric space $(E,d)$.
Thus, $E$ is dense in $E_{G}$, that is, it's a plenum (which means
that every open neighbourhood in $E_{G}$ of a Thing in $E$ contains
\emph{other} Things from $E$.) Thus, since we already had a physical
meaning for the time duration, \emph{this induces a physical meaning
for the spatial distance} $d$. Note that the metrical time $\tau$
and the distance $d$ are, again, one and fixed once and for all (that
is, we don't consider a given, fixed set of Things and the possibility
of having different metrics on it, since that's just a mathematical
device, which only occurs in the human mind, not in the Universe which
is just one and is the thing we are trying to model with all this.)
In this way, from the set of Things and their states we get a relational
theory of space and time with well defined physical meanings for time
durations and spatial distances, meanings which ultimately are only
intrinsically dependent on the physical interpretation of the Things
themselves (which is a primitive notion) and their states. 

$\;$

Thus, the standard notion of spacetime can be abstracted from the
set of all Things and their relations of separation and mutability
(note that we don't include change, which is not modeled by, e.g.,
General Relativity; instead, one derives conclusions from this theory
about the basic separation-mutability relations and, when living in
reality, one simply eventually experiences them but with change shaping
the actual way in which we do it\footnote{In fact, change seems to be rather tricky to define in a non-circular
or tautological way, since, no matter how one frames it, it always
tends to sound like something in the lines of ``change means that
something changes from this to that'', as if change were something
that can only be experienced.}, in particular, as a novel element of which the theory is silent,
since it can talk about processes, evolutions, etc., but always studies
them in the whole, the full $4-$dimensional manifold, field, the
full worldline, etc., the actual change is absent; this doesn't mean
that the theory says that change is ``unreal'', it simply never
entered into the things it describes in the first place; change may
be or may be not real, but General Relativity is not the theory that
will tell us which option is true; see the final paragraph of this
section for more about this.)

$\;$

The semantic formulation of General Relativity would be as follows:
at the mathematical side we have $(M,g_{ab};F_{a_{1}...a_{n}}^{(i)})$,
where $M$ is a $4-$dimensional smooth manifold (with the usual topological
requirements), $g_{ab}$ a smooth lorentzian metric, and $F_{a_{1}...a_{n}}^{(i)}$
a set of tensor fields, \emph{where the triple in question is a fixed
solution to the Einstein field equations} $G_{ab}=\kappa\sum_{i}T_{ab}^{(i)}$
(this in order to match the previous notion of a single and fixed
Universe), while, at the physical side, the set of all Things $\Theta$,
their set of \emph{legal} states $S_{L}$, any Thing $x$ with its
history (which can be considered to be an infinitesimal portion of
a physical field evolving in time\footnote{General Relativity is a field theory. A field as a whole is a single
thing instead of several, nevertheless, we \emph{postulate}, and can
think of it, as the total union of infinitesimal portions of the field
in different regions of spacetime and, in this way, we actually have
an uncountable infinity of things in this universe. What this actually
means is that the mere existence of a field of any kind gives rise,
\emph{via these axioms}, to the emergence of spacetime by the very
nature of the field as something infinitely divisible into minor things
or portions of it. The fact that actions take a while to go from one
portion to another is then implemented or reflected in the hyperbolic
nature of field equations in General Relativity. With this we can
also see more clearly why one has to choose a single solution to build
the spacetime. Consider, for simplicity, that a minimal field portion
is a field portion of the whole field on a small compact region $U$
of spacetime and build the phase space of this portion. If, instead
of considering the state of this portion only for a given solution,
we consider ``the state of the field'', then we would need to consider
the whole of that phase space, which, for example, containts an infinitude
of field configurations whose support is the compact region $U$.
But, of course, we want just one state per one of these small compact
regions (imagine, for simplicity, that they are the points of the
spacetime), since those states will be the points of the relational
spacetime which have to coincide, for each solution, with the manifold
to which the region $U$ belongs (see the paragraph in this section
about the ``problem of time'' for an additional clarification and
a more precise mathematical formulation.) This and the desire to fit
a single Universe is what motivates the assumption in question, that
is, the Universe is not a ``phase space'' which encompasses all
possible solutions; instead, only one of these solutions can describe
the single Universe in which we live.}) and proper duration $\tau_{x}$, and the different simultaneity
spaces $E$. The semantic map $\varphi$ takes $p\in M$ to $x\mid_{s}$
(that is, Thing $x$ in state $s\in S_{L}$; or, more precisely, \emph{just}
the state $s$ itself of Thing $x$; see next paragraph), $\varphi(p)=x\mid_{s}$,
a timelike curve $\gamma$ to the history $h$ of Thing $x$, $\varphi(\gamma)=h$,
the number $\tau\left[\gamma\right]\doteq\int_{\gamma}\sqrt{-g_{ab}\left(\frac{\mathrm{d}}{\mathrm{d}\lambda}\right)^{a}\left(\frac{\mathrm{d}}{\mathrm{d}\lambda}\right)^{b}}\mathrm{d}\lambda$
to the proper duration $\tau_{x}$ of $h$, $\varphi(\tau\left[\gamma\right])=\tau_{x}$,
and the distance of the riemannian metric $E_{ab}$ induced by $g_{ab}$
on $\varphi^{-1}(E)$ to $d$. Thus, General Relativity is an exactification
of our relational theory of spacetime \cite{key-1}.

$\;$

So, how do we build explicitly this spacetime in terms of fields,
then? In general, one just takes the Gravitational Field for this.
Indeed, one uses a coordinate chart to characterize the points in
$M$ (when there's an interpretation, we will do a slight notational
abuse and write $M$ instead of the correct $\varphi\left[M\right]$),
and then the proper times and distances of the metric $g_{ab}$ to
give physical meaning to these coordinates (the proper time of a timelike
curve is taken as its parameter and this parameter is taken as a time
coordinate, and the same with spacelike curves, distances, and spatial
coordinates; note how it's a property of the Gravitational Field,
the proper time, the thing that allows us to distinguish different
states separated by the temporal partial order via the different values
that this property takes on these states); but, since the metric is
actually a \emph{property} of the Gravitational Field (which is, of
course a matter field), what one is doing with this process is to
characterize the points $p$ via the values of properties of this
field, i.e., via states (of portions) of the latter (and, actually,
properties for which we have explicit physical interpretations.) Note
that a key property of spacetime is that the solutions for all fields
are fixed and that all fields \emph{superimpose} with each other \emph{on
the same point} of spacetime for all points (since they permeate all
of space), and this is modeled by the \emph{interposition operation}
$\dot{\times}$ at the ontological level; thus, we also \emph{postulate}
in our set of axioms that a point $p\in M$ is actually given by certain
\emph{equivalence classes} $p\doteq\left[s_{F^{(i)}}\right]$ of field
states (once spacetime emerges, the equivalence manifests itself as
the standard superposition principle of fields.) This gives us liberty
for choosing any of the fields to characterize space and time via
their states, and this is the \emph{origin} of the notion of \emph{spacetime
as a manifold}, i.e., a set of points which can be mapped to $\mathbb{R}^{n}$
in different, though relatable, ways; of course, this is reflected
in the liberty for choosing different coordinates systems, which are
nothing more than mathematical abstractions from the fact that a given
spacetime point can be characterized, in different but equivalent
ways, by the values of the properties of different fields (and how
many values per point is what will determine the dimension of the
spacetime; of course, we postulate that this dimension is equal to
$4$); this, in turns, leads us to \emph{postulate} the principle
of physical (this when solutions are fixed; more on this below) general
covariance, which states that the equations, and physical information,
of the fields cannot depend on the choice of a particular coordinate
system, thing which leads us to formulate them with tensor fields
and other intrinsic geometric objects. Nevertheless, note that fields
do not ``inhabit'' spacetime: \emph{they produce it}. Fields exist
by themselves, in an autonomous and absolute fashion, even when, for
convenience, one represents them mathematically as functions on spacetime\footnote{When we build the spacetime using a field, we are actually \emph{postulating}
that each minimal sub-Thing, of the infinitude whose aggregation forms
the whole field, is allocated at \emph{different} spacetime points,
and this implies the standard notion of the field as an extended entity
(and also separates points in $M$) in spacetime. Now, since all portions
have analogous properties (being the same type of entities), then
we can consider the same property (say, the intensity) and see how
it varies when we pass from each minimnal portion to another. Thus,
this is why (once we choose, as just mentioned, another one of these
fields and its states as a reference field for characterizing the
spacetime points) we represent fields as functions on $M$ (since
the two fields will generally have different values for their properties
on the same spacetime point on which their states superimpose, and,
therefore, if we use the values of the properties of one of the fields
as coordinates for that point, the values of the properties of the
other field will be represented as functions of those coordinates,
since the latter also characterize the state of this second field,
but by values other than the ones of its properties on that state;
this is also the case for the metric, since the values of the field
$g_{\mu\nu}(t,\overrightarrow{x})$ do not directly represent proper
times and distances.)}, this is just a mathematical representation that one shouldn't take
literally and translate it to the physics in such a naive way, as
if mathematics were a perfect and transparent mirror of physical reality:
it's not, since, as we saw, one can consistently build a relational
proto-physical theory of spacetime starting from Things. Note that
the functions of the form $f:M\,\longrightarrow\mathbb{R}$ are properties
of minimal portions of the field, since at each state $M\ni p\doteq\left[s_{F^{(i)}}\right]=(t,\overrightarrow{x})$
we get $f(t,\overrightarrow{x})\in\mathbb{R}$, that is, associated
to the state we get a real number which is just the values of the
properties in that state (but note that $f$ encompasses properties
of \emph{all} the minimal field portions, since its domain includes
all of space; for the portion that evolves along the curve given by
$\overrightarrow{x}=\overrightarrow{x}_{0}=const.$, then $f(t,\overrightarrow{x}_{0})$
indeed represents the value of a property for \emph{that} portion
as the latter evolves); due to general covariance, the point $p\in M$
could be constructed using another field, and, therefore, the \emph{same}
function $f$ \emph{also} represents properties of portions of this
other field; thus, the algebra $C^{\infty}(M)$ can be seen as the
algebra of properties of \emph{all} the minimal field portions of
\emph{any} of the matter fields (bear in mind, though, that this algebra
is \emph{not }the algebra of functions on the \emph{phase space} of
the \emph{whole} field, where each function there actually represents
a property of the whole field and also their domain comprises \emph{all}
the possible solutions.)

$\;$

Note that the notion of a given fixed spacetime $(M,g_{ab})$ but
with a matter field $F_{a_{1}...a_{n}}$ on it, considered as a variable
for its matter field equation and for which one studies the different
solutions (all over the same $(M,g_{ab}))$, is not something to which
we can give a physical interpretation, since it only exists in the
mind (the Universe is a single one and fixed) and is just a mathematical
device for studying the matter fields at different sub-regions of
the whole spacetime (since, in that case, the solutions can indeed
be different in each patch; these solutions would need to approximate
the actual full solution, particularized to each patch, of the Einstein
equations for the Universe) and when the interaction between gravity
and the rest of the fields can be ignored. 

$\;$

In retrospective, the only thing that deserves to be called spacetime,
and \emph{as a structure common to all the matter fields}, is just
the set $M\ni p\doteq\left[s_{F^{(i)}}\right]$ and its dimension,
that is, the bare differentiable manifold. Distances, durations, temporal
partial order, causality, and the distinction between mutability and
separation, are simply mere properties of the Gravitational Field.
Nevertheless, these properties result useful for characterizing $M$,
since one can easily associate them to curves, have physical interpretations
which in experience are perceived in a direct and obvious way by us,
and the fact that it's possible to consider a fixed Gravitational
Field as ``background'' to study the dynamics of the other matter
fields (when the backreaction can be ignored), while the converse
seems more complex since it doesn't seem to be very clear how one
is supposed to choose a solution for the other matter fields as background
due to the now variable metric being part of the non-gravitational
field's equations themselves (see next paragraph for more on this
point); as we mentioned before, this is actually the only ``asymmetry''
that remains between gravity and the other fields. Note that in our
relationist theory of how spacetime emerges from Things we used the
metric time to characterize space (both for the notions of separation
and metric distance), but this is legit since, due to the fact that
we are considering the real spacetime which is one and fixed, all
the solutions and notions are fixed and then we can use one aspect
from one to characterize another notion or viceversa (in this case,
since we already have physical interpretations for the states in the
Thing's history, then it's natural to introduce the partial order
since, because of this, it naturally, in a transitive fashion, acquires
a physical interpretation and so on.)

$\;$

The case in which both the spacetime metric $g_{ab}$ and the other
matter fields are variable and unspecified is even less pasible of
a full physical interpretation, and the manifold $M$ becomes just
an arbitrary abstract thing (the coordinates too, since one cannot
interpret them in terms of proper times and distances of the variable
metric because one uses those coordinates to study the very variability
of the metric via the field equations), whose only purpose is to remind
us of the fact that, whatever the solution is, it must imply mutability
and separation; this gives rise to the so-called ``Problem of Time''
in canonical quantum (but also in classical) gravity. We cannot associate
this situation to the real spacetime, which is one and fixed, it's
just a (very useful) mathematical fantasy that only exists in the
mind. The manifold and its coordinates remain here as residues, product
of abstracting them from the real ontology but for then artificially
give them an independent status from this ontology in our mathematical
models. Note that we defined spacetime as $M\ni p\doteq\left[s_{F^{(i)}}\right]$,
i.e., mathematically, it's the manifold \emph{and }a solution for
the fields $F^{(i)}$. The standalone, ``abstract'' manifold, and,
in particular, on which many different field solutions are considered,
is a mathematical fantasy and it doesn't mean a `` single and fixed
real spacetime'' on which these different solutions live (of course,
assuming the case of all fields variable, not the case with a background,
fixed metric.) Instead, it means that the mere mathematical structure
of the \emph{real} spacetime associated to each of those solutions
is the same for all of the latter (in particular, that's why one can
write two different of these solutions on the same ``abstract''
manifold, but where each solution must have different property values
for the same ``abstract'' manifold point, since it's the same ``abstract''
manifold structure but for different field configurations and real
spacetime, and therefore different physical field values are associated
to a same ``abstract'' manifold point, otherwise they would both
be just the same solution and real spacetime.) In particular, this
is why, in the definition of a spacetime singularity, the points of
the manifold on which the curvature diverges are not considered to
be part of the spacetime, since no Gravitational Field can be defined
on them, and, therefore, they would be points of a mere mathematical,
``abstract'' manifold. In a more precise way, consider the (kinematical)
phase space

\[
X=\left\{ \left[h_{ab},\pi^{ab}\right]\diagup h_{ab},\pi^{ab}\in C^{\infty}(\varSigma)\,(\mathrm{as\,fields}\,\mathrm{on}\,\varSigma)\right\} 
\]
of GR (where $h_{ab}$ and $\pi^{ab}$ are, respectively, a smooth
riemannian metric on a Cauchy surface $\Sigma$ in a \emph{compact,
boundaryless} $M$ and its conjugate momentum tensor density) and
the set of properties $C(X)$ of the field (note that these are properties
of the field on the whole spacetime and that their domain contains
points from any solution, these are \emph{phase space properties}.)
We start with space, since time will require more considerations.
Consider now the subset $\mathcal{F}\subset C(X)$, of properties
which are of the form

\[
F_{f}\left(\left[h,\pi\right]\right)\doteq\int_{\varSigma}f\boldsymbol{\epsilon}(h),\,\forall\left[h,\pi\right]\in X,
\]
where $f\in C^{\infty}(\varSigma)$ and $\boldsymbol{\epsilon}(h)$
is the volume element of $h$, i.e., $\boldsymbol{\epsilon}(h)=\sqrt{h}\,\mathrm{d}^{3}x$.
Evidently, for a field portion in a volume $V_{1}\subset\varSigma$,
the value of this property is

\[
F_{f}\left(\left[h,\pi\right]\right)=\int_{V_{1}}f(\overrightarrow{x})\,\sqrt{h}\,\mathrm{d}^{3}x,
\]
and for an infinitesimal field portion at $\overrightarrow{x}$, simply
\[
f(\overrightarrow{x})\left(\sqrt{h}\,\mathrm{d}^{3}x\right)\mid_{\overrightarrow{x}}
\]
(seen here as a $4-$form $f(\overrightarrow{x})\boldsymbol{\epsilon}(h)\mid_{\overrightarrow{x}}$,
so that it's actually mathematically well-defined; note, though, that
there's a 1-1 correspondece between $f(\overrightarrow{x})\left(\sqrt{h}\,\mathrm{d}^{3}x\right)\mid_{\overrightarrow{x}}$
and the actual value $f(\overrightarrow{x})$, since the functional
form of the volume element is known and fixed and the only arbitrary
variation is in the function $f$, and, in this way, we can consider
these latter finite values, the\emph{ volumetric densities} at $\overrightarrow{x}$
rather than the infinitesimal quantities.) Thus, for a given phase
space point $\left[h,\pi\right]$, the values of those properties
for an infinitesimal field portion at $\overrightarrow{x}$ are given
by the values of the functions $f(\overrightarrow{x})$; but, since
all the functions $f$ cover all the possible smooth functions on
$\varSigma$, which we know should indeed correspond to all the possible
properties of the infinitesimal field portions for the spatial part
of a solution, this means that $\overrightarrow{x}$ characterizes
the values for \emph{all} the possible properties of the infinitesimal
field portion, that is, the generic state $s$ of this portion is
identified with the generic space point $p\in\varSigma$ of coordinates
$\overrightarrow{x}$. In this way, if we relationally build space
via the states $s$ of infinitesimal gravitational field portions
using this phase space, we get, for \emph{any} solution, an identification
of relational space with the initial manifold $\varSigma$ that was
assumed to represent space once a solution is chosen (of course, $\varSigma$
is interpreted as the physical space \emph{only after} a given solution
is considered; note that the manifold is there all the time since
it was assumed from the onset that space would have that type of mathematical
structure and not because everything happens on a ``background''
space.) Furthermore, one can intuitively see that, if the phase space
functions vary over all possible phase space points, and if space
points are states of infinitesimal portions of the field for a given
solution, then a phase space property $F_{f}$ whose domain of integration
is restricted to \emph{all} the \emph{different} infinitesimal portions
of the field for a given solution will \emph{give rise} to a \emph{space
function} in the space algebra $\mathcal{A}_{sp.}=C^{\infty}(\varSigma)$
for the space that relationally arises from that solution. That is,
in this view, the space functions, rather than being physically independent,
actually come, or are derived, and acquire physical and mathematical
meaning from phase space functions (which are taken now as the fundamental
objects) via certain process once space relationally arises when a
solution is fixed (the value of the space function on an event is
actually the value of a property for an infinitesimal field portion
in the state identified with that point; of course, this may seem
trivial and that we put the space functions there by hand, but the
relevant thing here is the reversion in the logic: in the usual approach,
we have the space functions, which exist independently from the fields,
while, here, without the physical field, no space function can arise
in the first place.) Now, since, of course, we can consider more general
phase space functions too, for the Gravitational Field, length can
also be seen in this way (that is, as a property of infinitesimal
portions of the field, as claimed before), since it's a functional
of the form $l\left[\gamma;h_{ab}\right]\mid_{0}^{x_{1}}$ (with the
curve of the coordinate $x_{1}$ taken as $\gamma$), and therefore
\[
\widetilde{l}(\gamma)\left[h_{ab},\pi^{ab}\right](\overrightarrow{x})\doteq\sqrt{h_{11}(\overrightarrow{x})}\delta(x_{2},x_{3})\frac{1}{\sqrt{h(\overrightarrow{x})}}
\]
is an authentic (distributional) density, whose integral is the value
$l\left[\gamma;h_{ab}\right]\mid_{0}^{x_{1}}$ itself. This is why
General Relativity is special and much more revolutionary than what
it already seems at first: in its phase space formulation, and unlike
other non-gravitational theories, the manifold $M$ used there no
longer has physical meaning as spacetime, since \emph{all} fields
are \emph{unspecified} and therefore no \emph{a priori} solution to
any physical field can be used to relationally build the spacetime,
even the manifold $M$ itself.

$\;$

Thus, one should be very careful in making literal interpretations
of the mathematics. When there are fixed solutions, they indeed relate
to physical spacetime and the coordinates have physical interpretations,
since they ultimately refer to the values of the properties of the
matter fields (in these solutions.) But, in the variable metric and
other fields too situation, one loses this, the coordinates become
mere abstract mathematical parameters and the general covariance\footnote{The coordinates now are abstract residues, but the covariance remains
as a reminder that the solutions, once fixed, must be independent
from them (which for this case do have a physical interpretation.)} of the gravity field equations with respect to them now looks very
similar to the gauge symmetry in, e.g., electromagnetism, where $A'_{a}=A_{a}+\partial_{a}\chi$
(here, both $A'_{a}$ and $A_{a}$ give rise to the same $F_{ab}$,
since $F_{ab}\doteq\partial_{[a}A_{b]}=\partial_{[a}A'_{b]}$, with
$\chi$ being an abstract scalar field without physical interpretation.)
We refer to this second case as gauge general covariance, as opposed
to physical general covariance. In the case of a fixed metric with
variable non-gravitational matter fields (and no backreaction being
considered), we have physical general covariance of the non-gravitational
matter fields equations, but, since the coordinates have physical
interpretations thanks to the fixed metric solution, the invariance
of the non-gravitational matter field action under these transformations
doesn't generate constraints like in the case for the gravitational
action. That is, the link (invariance of the action)$\Longrightarrow$(constraints)
is broken here, due to the fact that the coordinates are not mere
abstract, gauge mathematical parameters for this situation, and, actually,
we can get non-trivial Hamiltonians/metric time evolutions. The mathematics
of the models indeed recognizes these differences, since, in the fixed
metric case, there's no implied intrinsic variability of the metric
from the part of the non-gravitational actions (the metric is just
an external thing to this action), while metric variability is, of
course, intrinsic in the gravitational action; the proof that this
implies, respectively, the validity or not of the link (invariance
of the action)$\Longrightarrow$(constraints) is shown explicitly
in the actual calculations, where in both cases the actions are invariant
under the coordinate changes, but only in the second case we get constraints
associated to these invariances. The Problem of Time could be partially
solved if one can make sense of the case with variable metric but
with a fixed non-gravitational matter field solution (and no backreaction
being considered), since one could study the gravitational variability
by parameterizing the points of the real spacetime $M$ with the values
of the properties of these other matter fields (although, of course,
one cannot interpret the resulting ``time'' as duration, since that's
a property of the Gravitational Field, and, instead, the interpretation
is as whatever the properties of these other fields are)\footnote{Although, this seems problematic, considering that the metric enters
into the equations of these other fields via the Levi-Civita connection,
that is, the equations of these other fields depend on the variable
metric. On the other hand, since the other fields don't enter into
the pure gravitational action, the latter still interprets that there
isn't any fixed spacetime and that's why there still are constraints
(compare with the case in which the other fields are variable and
the metric is fixed.) In general, the only way in which the other
fields relate to gravity is via the Einstein equations, and this is
done by adding the other field's actions to the pure gravitational
action; but, of course, this would imply the variability of the other
fields, since one would be considering their field equations via their
action. One could try to put these other fields into the Einstein
equations by hand, but, since they are fixed, this would restrict
the variability of the metric. Thus, all this seems to point out that
this other possibility (variable metric and other fields fixed) is
untenable or forbidden by the formalism itself.}. But, in the case with metric and other fields taken as variable,
the resulting formalism is pure mathematical fantasy, which, of course,
cannot have any time in it, since that formalism doesn't have any
relation to real spacetime (although, the formalism in question is
the only way we have to express the equations of gravity.) The only
possible intepretation in this case is in ``potential'' terms, that
is, if all the solutions maintain a same peculiarity (e.g., a given
hypersurface is spacelike for all the solutions, or the fact that
for any solution $M$ will refer to mutability and speration.) Nevertheless,
all these interpretations are only potential, in the sense that, eventualy,
once a solution is fixed, the actual intepretation holds; the formalism
itself has no true, actual physical interpretation in this sense.
Thus, in the phase space of Gravity, there's no background spacetime
(both in terms of manifold and metric), and this leads to a Hamiltonian
$\mathcal{H}_{EH}=\sum_{i}N_{(i)}\mathbf{C}^{(i)}$ given by the constraints
$\mathbf{C}^{(i)}$, which, in turn, leads to gauge invariant phase
space properties $F$ (i.e., the ones that commute with the constraints,
$\left\{ F,\mathbf{C}^{(i)}\right\} =0$) that lack an adequate time
evolution (since $\frac{\partial F}{\partial t}=\left\{ F,\mathcal{H}_{EH}\right\} =$$\sum_{i}N_{(i)}\left\{ F,\mathbf{C}^{(i)}\right\} =0$.)
The only solution for this problem is to consider a solution $g_{ab}^{1}$
to the gravitational equations, to particularize the gauge invariant
phase space function $F$ (whose domain comprises all possible solutions)
for $g_{ab}^{1}$, and then make the change of coordinates $t\,\longrightarrow\text{\ensuremath{\tau}}$
(where $\tau$ is the proper time according to $g_{ab}^{1}$) in order
to study the \emph{true} time evolution (with respect to $\tau$,
of course) of $F$ and of properties such as $q_{ab}^{1}(\tau)$,
the curvature, etc., and this process would need to be considered
as a separate case for each different solution; alternatively, one
can start with a solution $g_{ab}^{1}$, a chart defined in terms
of $\tau$, and then build the natural gauge invariant properties,
such as the curvature. In any case, the solution, and therefore the
physical spacetime, is always needed (note that, as mentioned before,
the field has to be the gravitational field and one cannot simply
consider any other field as fixed for doing this, since, due to the
special role of the gravitational field as the one that gives the
spacetime metric, if this field is taken as variable, then all other
fields must be taken as variable too.) Thus, to obtain an actual time,
one needs to fix a solution. In phase space, the metric is the intrinsic
variable, and therefore no solution is assumed, this is why the ``time''
evolution there is only gauge, and if one goes to the quantum theory
with this picture, this gives the impression that there's ``no time''
in quantum gravity. But this is just an artifact of the phase space
picture, and can be solved in the classical case in the way we mentioned
above. The real problem in quantum gravity is the one related to the
continuum. But if we don't have continuum, we don't know how to pass
to the spacetime picture, which in the classical case is something
as trivial as just picking a given solution, and then we are stuck
in the phase space picture with no time! This is the real core of
the problem of time once we are in quantum gravity. To solve it, one
needs an analogue of the spacetime picture, but, of course, without
the continuum. Note that a fully diffeomorphism invariant property
is not necessarily a gauge invariant phase space property, since the
symmetries implemented in phase space by the constraints do not exhaust
the full diffeomorphsim group of the manifold $M$; among those properties
are, for example, the spacetime volume; it's often argued that the
``real'' properties are the gauge invariant phase space ones; but,
again, this is another artifact of the phase space picture: those
other properties are certainly not gauge invariant phase space properties,
because the phase space picture is artificially restrictive, but make
perfect physical sense as properties in the spacetime picture (furthermore,
one doesn't calculate their time evolution via phase space, but, as
mentioned above, via the spacetime picture: time evolution is the
specific change with respect to a \emph{duration}.)

$\;$

In order to \emph{define a coordinate} and, in this way, to \emph{characterize}
in a concrete (and, of course, relational) way \emph{a point of the
``abstract'' manifold}, consider the \emph{metric area} of a surface
$S\subset\varSigma$, sub-manifold of $\varSigma$, that is, 
\[
\mathrm{a}^{S}\left(\left[h,\pi\right]\right)=\int_{S}\boldsymbol{\epsilon}(q),\,\forall\left[h,\pi\right]\in X
\]
(where $q$ is the metric induced in $S$ by the metric $h$ in $\Sigma$.)
This is done, for example, when using spherical coordinates in a metric
$h$ with spherical symmetry, where the radial coordinate of any point
$p$ in a topological sphere $S$ is defined as\footnote{To see it as a property of an infinitesimal field portion, i.e., as
a density, proceed as in the case of the length.} 
\[
r\left(p\right)\doteq\sqrt{\frac{1}{4\pi}\mathrm{a}^{S}\left(\left[h,\pi\right]\right)}
\]
 Now, as we move to bigger spheres, $r$, of course, just monotonically
increases, and, in this way, this property of the field separates
the points in $\varSigma$ and gives them \emph{physical} meaning,
since it assigns to each of them a different value (along a curve
of fixed angular coordinates, of course; note, too, that for being
able to do this, a manifold structure on $\varSigma$ \emph{is assumed
in the first place}.) This process is indeed the \emph{relational
way} we use everyday in GR to define the points of the manifold (since
we use the values of properties of the Gravitational Field to characterize
them), but it's too tied to the classical structure of the theory
for being of any general use.

$\;$

Indeed, consider the quantization of the phase space property $\mathrm{a}^{S}$
in LQG. There, the differentiable manifold $\varSigma$ is maintained,
and $\mathrm{a}^{S}$ is promoted to an area operator, $\hat{\mathrm{a}}^{S}$.
The eigenstates of this operator are the so-called spin network functions,
the simplest of which is the Wilson loop of spin $s\in\frac{1}{2}\mathbb{N}\bigcup\left\{ 0\right\} $.
If this loop punctures the surface $S$, then its area eigenvalue
is proportional to $\sqrt{s(s+1)}$. If we now consider different
spheres, as in the classical case, then the area eigenvalue will still
always be the \emph{same} if the loop still punctures these other
spheres too. But, since, in a relational approach, these values give
physical meaning to the points of space, this means that all these
different spheres in $\varSigma$ are physically the same. Thus, there's
an obvious conflict between the physical meaning of this operator
and the surfaces to which it supposedly gives their area. In the most
benign case, this may indicate that this particular mathematical incarnation
of the formalism is just very awkward, but in the worst case, that
it's actually a wrong approach. The more sensible position seems to
be the latter one, since, as was mentioned, the manifold is a direct
consequence of the assumed continuous range of classical properties.
Thus, one cannot just dispose one and keep the other, since the \emph{coordinate
charts are just a mathematical abstraction of the values of field
properties}; if the values of field properties are discrete, then
no classical charts and, therefore, no classical manifold. In this
way, since this formalism contradicts itself from the physical point
of view, one should completely reformulate the approach in a way that
avoids this type of problems.

$\;$

In addition, and from a relational point of view, there's also a second
issue with the phase space functionals for geometrical properties,
like the previous $\mathrm{a}^{S}$, which is the following. In a
relational theory, \emph{only} after a solution $g$ is selected does
the manifold $\varSigma$ (and the surface $S)$ acquire physical
meaning. This means that only the spacetime picture version, $\mathrm{a}^{S}\left(\left[h,\pi\right]\right)$,
of the phase space functional $\mathrm{a}^{S}$ can be given the interpretation
of area of $S$, since in that case $S$ has a physical interpretation
as a surface in physical space, while the phase space \emph{functional}
cannot be given the mentioned interpretation because no solution is
selected, and, therefore, the $S$ in it is just an \emph{abstract
parameter}. In this way, if we go to the quantum theory, the eigenvalues
of $\hat{\mathrm{a}}^{S}$ cannot be \emph{strictly} interpreted as
the possible area values for a \emph{physical} surface $S$ (even
when one suspects that this spectrum is on the right track.) Furthermore,
if we simply eliminate the continuum from the phase space area functional,
we lose all the topological and differential geometric information,
which was encoded in $S$: we would be calculating the area of something
which is missing! \emph{Thus, we conclude from this that a spacetime
picture is needed if one pretends to study geometrical properties
like areas and volumes: it's only when both metric, $q$, and surface,
$S$, meet that they acquire, simultaneously, physical meanings, which
are also mutually consistent with each other.}

$\;$

Finally, it's usually said that something exists if it exists at an
instant of time in the present. In the context of our ontology, this
is misleading. Reality is a single thing and a Thing in it either
exists or not, independently of time. If a Thing indeed exists in
Reality, then it has a state space with several states accessible
to it. Now, as the Thing changes, it's actually its state the one
that changes. That is, the thing that disappears, so to speak, as
the Thing changes, is the previous state, not the Thing itself. These
are two levels of concepts that shouldn't be confounded (if a given
Thing is annihilated, the basic ontological substance is still conserved
and a new Thing of a different kind has to emerge, like photons emerge
from the annihilation of an electron with a positron.) What presents
itself and unfolds in time or change is the Thing \emph{in} a particular
state. Spacetime is the network of spatiotemporal relations between
all the states of all Things in Reality. Regarding the relativity
of simultaneity in special relativity, the changes that this introduces
in the pre-relativistic view on space, time and change are the following.
Consider the worldline $\gamma_{A}$ of a changing Thing $A$. As
the Thing changes, we reach the event $e\in\gamma_{A}$ at the present
moment. In pre-relativistic physics, one can establish what are all
the events that are simultaneous with $e$, what events are in the
past of this absolute surface of simultaneity and which ones are in
its future. This exhausts all the events in spacetime. Events in the
past are taken as non-existent anymore, events in the simultaneous
present as currently existing, and future events as still not existing.
In this way, as the Thing changes, the surface of simultaneity advances.
This view is, of course, that of presentism, and is the natural one
if one accepts change as something real. In special relativity, one
can only establish the absolute causal past $J^{-}(e)$ of event $e$
and its absolute causal future $J^{+}(e)$, but no notion of surface
of simultaneity, to which $e$ would belong, can be determined objectively,
the only present for event $e$ is just the event itself. Nevertheless,
one can establish that events in $J^{-}(e)$ already happened, and
not only according to Thing $A$, but also according to any other
Thing. For example, if $\gamma_{B}$ is the worldline of changing
Thing $B$ and event $p\in\gamma_{B}$ is also in $J^{-}(e)$, then
we can affirm that, according to $B$, event $p$ already happened
to it (since, once $p$ is in $J^{-}(e)$, if $B$ emitted a light
signal at $p$, this signal hits $A$ in $e$, but, then, when $A$
receives, it concludes that $B$ already left event $p$ behind.)
Symmetric notions apply to the causal future. Thus, the causal past
and future are objective and absolute: what happened, happened. The
so-called block view of spacetime, in which all events somehow always
exist, is not necessary. Indeed, it would be necessary if, say, event
$e$ that, according to $A$, is present, is nevertheless in the past
of, say, $B$ (assuming that the spacetime is causally well-behaved);
but, if this were the case, this would be contradictory, since, then,
$B$ could be receiving, in its present, light signals that $A$,
according to its present, has not even emitted yet. This contradiction
only dissolves if the present of $B$ is neither in $J^{+}(e)$ nor
$J^{-}(e)$, but then the block view is not necessary anymore. In
this way, for each changing Thing, one can establish a present, composed
of a single event in its worldline, and an absolute causal past and
future, for which the events in the former are not anymore and the
events in the latter still not are, and where any other changing Thing,
according to its own present, will agree that those events indeed
are not anymore or still not are. Thus, what we get is a picture vaguely
similar to that of presentism, in the sense of becoming and absolute
(causal) pasts and futures, but where the notion of surface of present
is not valid anymore. In the case of a field, the changing Thing is
any portion of the field localized along any timelike worldline (this
is also valid for the metric tensor field itself once it's fixed,
since in that case one can take the proper time of the worldlines
as coordinates, while this is not possible if the metric is considered
as the variable, since the whole formalism doesn't even correspond
to the physical spacetime, as we mentioned before.)

$\;$

\subsection*{{\normalsize{}1.2 Critique of the Theory in 1.1}}

$\;$

While the theory indeed success in giving a purely relational proto-physical
theory of spacetime, it has a particularly ad-hoc flavor. Pretty much
everything is postulated and the theory, at best, shows that all these
suppositions can hold together to give the desired theory, without
needing to assume another substance besides matter, aspect which captures
the relationalism one is looking for. Let's recapitulate all the ad-hoc
assumptions: 1) Mutability, which basically amounts to postulate as
axiom the \emph{very essence} of time; 2) the temporal partial ordering,
which gives rise to durations; 3) the finite speed propagation of
causal influences, which gives rise to distance and Separation, the
latter being, again, akin to postulate as axiom the \emph{very essence}
of space; 4) the field's nature as an infinitely divisible and extended
Thing; 5) the (differentiable) manifold structure of the set $M$
and its dimensionality. We will see in the next sections how the mathematical
formalism of the spectral standard model can be used to make a proto-physical
relational theory of spacetime in which all of the mentioned points
1),...,5) can actually be derived rather than pressuposed.

$\;$

Note that the mathematical model of the spacetime as a manifold was
actually already included in the proto-physical theory that was built
in the previous section, and this was a consequence of the particular
assumptions that were made in that theory. Thus, the distinction between
$M$ and $\varphi\left[M\right]$ in the semantic interpretation of
the mathematical formalism of General Relativity which we did is more
a mere book-keeping device. The true non-trivial content of that semantic
interpretation resides in the metric $g_{ab}$, which is a particular
mathematical construct, being interpreted in terms of the durations
and distances of the relational proto-physical theory, since those
physical notions in the latter didn't have a precise \emph{quantitative}
meaning, and, therefore, what the semantic interpretation introduces
is an actual way for calculating those values, and this will determine
a range of consequenes, proper of this particular physical theory.
Note also that, in this interpretation, the functions in $C^{\infty}(M)$
acquire their physical interpretation from that of $M$ in a transitive
way.

$\;$

But, consider now the following situation. What if we assume the whole
mathematical formalism of General Relativity and \emph{only} give
a physical interpretation to $C^{\infty}(M)$ and as the algebra of
properties of \emph{all} the minimal field portions of \emph{any}
of the matter fields? If we do this, we can take a point $p$ from
the (abstract) set $M$ and consider the values $f(p)$ for some $f\in C^{\infty}(M)$.
But, since those are values of properties, then this implies that
the points $p$ must be the states of those minimal portions, that
is, they acquire this interpretation transitively from that of $f$
as properties. The assumption in the mathematical formalism about
$M$ being a set with different points then implies Mutability, and
its manifold structure implies the other assumptions made in the original
proto-physical theory. Indeed, the fields being represented as functions
on $M$ implies now the infinite subdivision property that they have.
The abstract metric $g_{ab}$ induces a partial order in the histories
made from points in $M$, which, given their interpretation as states,
is a temporal order and therefore implies physical duration (and also
exactified.) The hyperbolic nature of the field's equations implies
the finite speed (and a maximal one, not matter the specifics of the
influence) of causal influences, which, in turns implies the physical
Separation of the points in $M$ and implies that the mathematical
distance calculated from the metric is the actual physical distance
(and, again, also exactified.) Thus, we recover the previous proto-physical
theory of spacetime, where now the mathematical assumptions made in
the mathematical formalism become the physical or ontological assumptions
made in the ontological theory discussed initially.

$\;$

From the semantical point of view, these two different ways are certainly
equivalent, and also note that the ``input'' information needed
is basically the same, since, what was an ontological assumption in
the first approach, now must be assumed in the mathematical formalism,
and, thus, there's really no gain in terms of explanatory power by
shifting from one approach to the other. The advantage of the first
approach is that we can concentrate in the actual aspects concerning
to physical spacetime without making any compromise with a particular
exactification, and, then, this adds philosophical clarity to the
discussion and also means we can use this proto-physical theory to
interpret physically other physical theories as well (say, the space
and time notions in classical mechanics.) On the other hand, the advantage
of the second approach is that the ontological theory of spacetime
is now suggested or derived from the mathematical formalism of a given
physical theory, and, since science evolves and corrects itself, this
seems more convenient because the ontological theory will be up-to-date
with the latest developments, while the first approach risks at getting
trapped in assumptions that are later discarded by more up-to-date
physical theories. Thus, the adequate way of proceeding is actually
the following. One should take a given physical theory and study the
implied ontology, and then abstract the latter in a separate proto-physical
theory in order to achieve more philosophical clarity. Evidently,
the relational proto-physical theory of spacetime discussed in the
previous section was abstracted from General Relativity in its current
differentiable manifold mathematical formulation, in which most of
the intuitive notions we associate to space and time are put in by
hand into that theory, since they do seem to be fundamental, at least
at the level to which this physical theory operates.

$\;$

\subsection*{{\normalsize{}2.1 Spacetime Relationalism and the Spectral Standard
Model of NCG: The Spectral Standard Model and NCG}}

$\;$

Consider a compact, $n-$dimensional spin manifold $M$ and its associated
Dirac operator $\cancel{D}$ (which is the covariant derivative on
sections of the spinor bundle, i.e., the spinor fields.) It's known
that such spinorial structure gives rise to a riemannian metric $g_{ab}$.
Using the volume element $\boldsymbol{\mathbf{\boldsymbol{\varepsilon}}}=\sqrt{g}\boldsymbol{\epsilon}$
of this metric, we can build the space $L^{2}(S_{n})$ of square integrable
spinor fields and in this way consider the \emph{canonical spectral
triple}, defined as $(L^{2}(S_{n}),C^{\infty}(M),\cancel{D})$. One
can also show there that $\mathrm{tr}^{+}\,(f\mid\cancel{D}\mid^{-n})$$=k_{n}\int_{M}f\boldsymbol{\mathbf{\boldsymbol{\varepsilon}}}$,
where $f\in C^{\infty}(M)$ and $k_{n}$ is a constant. Then a \emph{general
spectral triple} is defined as $(\mathcal{H},A,D)$, where $\mathcal{H}$
is a Hilbert space, $A$ is a sub $^{*}-$algebra of bounded operators
(not necessarily commutative), and $D$ a self-adjoint operator that
mimics some properties of $\cancel{D}$; the\emph{ general or ``non-commutative''
integral} is defined as $\fintop a\doteq\frac{1}{k_{n}}\mathrm{tr}^{+}\,(a\mid D\mid^{-p})$,
where $a\in A$ and $p$ is certain number. The seminal\emph{ reconstruction
theorem of NCG }states that if a general spectral triple satisfies
certain regularity assumptions and $A$ is \emph{commutative}, then
there's a compact, $n-$dimensional spin manifold $M$ such that $(\mathcal{H},A,D)\cong(L^{2}(S_{n}),C^{\infty}(M),\cancel{D})$
(and, of course, this is what justifies to see general spectral triples
as genuine generalizations of the notion of spin manifolds to the
non-commutative case and to call $\fintop$ a genuine generalization
of the notion of metric integration to the non-commutative case) \cite{key-3}.

$\;$

The key observation on which the spectral standard model is based
is the following \cite{key-3}. For a canonical spectral triple, one
gets $\fintop fD^{-2}=k'\int_{M}fR\sqrt{g}\boldsymbol{\epsilon}$,
where $R$ is the Ricci scalar, that is, the standard Einstein-Hilbert
action. Now consider an \emph{almost commutative} spectral triple
based on the algebra $C^{\infty}(M)\otimes M_{N}(\mathbb{C})$ (where
$M_{N}(\mathbb{C})$ are the $N\times N$ complex matrices), Hilbert
space $L^{2}(S_{n})\otimes M_{N}(\mathbb{C})$ (the latter equipped
with the Hilbert-Schmidt norm), and Dirac operator $D=\cancel{D}\otimes I$.
Then the so-called \emph{spectral action} given by $\mathrm{tr}^{+}\,(F(D/\varLambda))$
(where $F$ is a real, positive even function), which is purely geometric
and generalizes to the non-commutative case an action based on a function
$F(K)$ (where $K$ is a scalar of the curvature), gives rise to the
standard action of gravity \emph{coupled to a Yang-Mills field}. That
is, we get, via non-commutative geometry, a ``purely geometric''
\emph{unification} of the forces. The metric is, of course, given
by the Dirac operator $\cancel{D}$, while the gauge fields correspond
to ``fluctuations of the metric'', and the gauge symmetries come
from the group of automorphisms of the almost commutative manifold
(in the commutative case, this group is just the group of diffeomorphisms
of the manifold.)

$\;$

The fact that the spectral action on an almost commutative space allows
one to derive the standard gravitational equations as well as the
existence of Yang-Mills fields and their equations, all this from
a non-commutative version of a purely geometrical action means that
the ``asymmetry'' between gravity and the rest of matter fields
that the spacetime substantivalist could claim is \emph{completely
erased} now, and, therefore, this provides a strong argument for substantial
monism in physics. Of course, this substance, rather than being ``all
geometry'' is just ``all matter'', and the previous unification
of the equations means that the dynamics of \emph{all} the matter
fields can be expressed in ``geometric'' terms, which simply means
that all of geometry, both the commutative part as well as the non-commutative
one, is nothing more than just a substantial property of these matter
fields. The Gravitational Field is only ``special'' in the fact
that the geometry of the usual spacetime $M$ is one of its properties,
while the geometrical elements which are a property of the other fields
(the fluctuations of the metric) correspond to phenomena that arise
due to the finite dimensional algebra part. But this is not surprising
since, being geometry always a property of matter fields and being
the usual spacetime commutative algebra a part of the tensor product,
then its geometry will necessary have to be associated to \emph{one}
of the matter fields. Besides, one could say that there's truly just
\emph{one} matter entity, since the spectral action is a single one
and referred to the algebra of the whole almost commutative spacetime;
thus, the rewriting of this as fields \emph{on} $M$ (the standard
commutative spacetime) and satisfying their particular field equations
can be seen as a mere convenient way of expressing the total information
of this single entity and that shouldn't be taken very literally in
a ontological sense.

$\;$

\subsection*{{\normalsize{}2.2 A Relational Theory of Spacetime from the Spectral
Standard Model}}

$\;$

Here we do something which was already partially explained in section
1.2. We formulate an apparently ``new'' physical theory by the following
axioms. At the mathematical side, assume a commutative spectral triple
$(\mathcal{H},A,D)$ that satisfies the regularity conditions of the
reconstruction theorem. We now interpret physically the algebra $A$
as the algebra of properties of \emph{all} the minimal Things (where
the algebra elements for some different minimal Things may be the
same), and that the dynamics of these Things is constrained by the
spectral action. If we now apply the reconstruction theorem, we get
that $(\mathcal{H},A,D)\cong(L^{2}(S_{n}),C^{\infty}(M),\cancel{D})$
and $\fintop a=\int_{M}f_{a}\boldsymbol{\mathbf{\boldsymbol{\varepsilon}}}$. 

$\;$

In the usual relational construction, the states $s$ are such that
the properties $f$ take a value $f(s)$ in that state. That is, the
properties are \emph{functions} of the state and of the state space
$M\ni s$, i.e., $f:M\,\longrightarrow\mathbb{C}$. This scheme \emph{doesn't}
say \emph{how many} states are, and, therefore, in order to have Mutability
and Separation, one must \emph{postulate} that there are at least
two states in $M$. This because the set $M$ is taken as ontologically
fundamental and the properties \emph{refer} to it for their mathematical
definition and implementation, and this means that the structure of
$M$ (including its cardinality) always will be fundamental ontological
postulates. Nevertheless, even if there's only one single state $s\in M$,
the very definition of property as a function $f:M\,\longrightarrow\mathbb{C}$
already implies the existence of an infinite number of different properties,
since there are as many different of those functions as elements in
$\mathbb{C}$. Note that this is not an exclusive consequence of taking
$\mathbb{C}$ as the quantity-value object, since this just makes
that the result is at least a possibility (since, if that set had
one single element, then there would be only one single possible function);
what makes the result to follow is the definition of the property
as a function and the axioms (ZFC) of Set Theory (we will investigate
this further in the next section.) Now, by the Gelfand duality, each
$s\in M$ has associated to it an algebraic state given by a Dirac
measure $\delta_{s}:C(M)\,\longrightarrow\mathbb{C},$ $\delta_{s}(f)\doteq f(s),$
$\forall f\in C(M)$. Thus, $M$ is identified with the set of pure
states on the algebra. Now, one could take the algebra and its structure
as fundamental and the states as the ones that \emph{refer} to it
for their mathematical definition and implementation. Furthermore,
these algebraic states $\omega:C(M)\,\longrightarrow\mathbb{C}$ also
associate values to the algebra elements via the expectation value
$\omega(f)$, which, for the previous pure states is just $f(s)$;
thus they indeed generalize the notion of state that was define only
in the context of the ``states-first'' approach via $M$ and the
functions on it as properties. In this way, the Mutability and Separation
in $M$ translates to the existence of at least two of these algebraic
states and these notions are now also valid for the case of a non-commutative
algebra $\mathcal{A}$, since the notion of algebraic states still
makes sense in that context. The set of pure states will only be a
manifold in the case of a commutative algebra. This approach is, of
course, more natural in quantum theories, where the very formalism
of that theory is a non-commutative probability theory which replaces
the classical probability theory given by a measure on phase space.
Also, since now the states are the functions of the form $\omega:\mathcal{A}\,\longrightarrow\mathbb{C}$,
even if the algebra has one single element, the existence of Mutability
and Separation are now a consequence of $\mathbb{C}$ having more
than one element, the definition of states as functions and the axioms
of Set Theory, by a reasoning identical to the previous one. Thus,
the change to this algebraic view shows that $\mathbb{C}$ having
more than one element, the definition of states as functions and the
axioms of Set Theory are a \emph{suficient} condition for obtaining
Mutability and Separation in the implied ontological theory (in the
scheme with $M$, these are \emph{necessary} conditions, but \emph{not}
suficient ones, since the cardinality of $M$ must still be postulated;
note that we are analyzing this only at the level of the most basic
structure, namely, Set Theory and functions, and, therefore, we don't
postulate a differentiable manifold structure on $M$ at this point.) 

$\;$

The Dirac operator allows to reconstruct the metric $g_{ab}$, while
the spectral action implies the vacuum Einstein field equations for
it (the metric is, of course, of riemannian signature, and, therefore,
we will have to add as an axiom that what we actually recover here
is the action in its Wick rotated form of an originally lorentzian
theory; we must do this since we need the hyperbolic character of
the equations to recover space; of course, it would be more desirable
to have a theory of lorentzian spectral triples and to prove the emergence
of the standard model from them.) Thus, the rest of the argument goes
as in the second approach to semantic interpretations discussed in
section 1.2, and we simply recover General Relativity and its implied
relational ontological theory of space and time. Note that the reconstruction
theorem alone is not enough, since it would only give the Mutability;
we also need the spectral action and therefore the field equations
in order to give physical meaning to a part of $M$ as describing
physical Separation. 

$\;$

Note that the previous construction, since it uses the algebraic view,
is such that the basic object is the one of properties, while the
notions of states sits on it, and this allows us to build a theory
which includes the notions of properties, states, and symmetries over
the basic notion of properties. This is desirable in a relational
approach to spacetime, since, given that the basic definition of material
Thing is as a substantial element with properties (states are not
mentioned), if only Things exist, then is natural to build the mathematical
formalism which will model Things as one based on notions that are
actually fundamental to Things (that is, their properties), and not
derived notions, like states.

$\;$

One reason why this approach is relevant is that it explains more
than the usual approach because the assumptions made now are all of
a same type (the ZFC and restrictions and regularity assumptions on
the structure of the spectral triple, where the latter models the
structure of the properties of minimal Things), while in the old approach
seems like a set of disconnected and arbitrary assumptions of different
kinds, thing which gives it the discussed ad-hoc flavor; thus, since,
as mentioned, the assumptions on the spectral triple have nothing
to do, in principle, with the standard, intuitive notions of space
and time, which are ad-hoc in the usual approach, we could say then
that our ``new'' physical theory here, based on the spectral standard
model, truly \emph{explains} in a non trivial/ad-hoc way the origin
of these notions from Things and their properties in a relational
way. It's also relevant because in this second approach it's made
manifest a supposition, the commutativity of $A$, which in the first
approach cannot be relaxed while in the second it can (but, in spite
of this, the notion of metric integration still makes sense.) Thus,
even if we start only from minimal Things and their properties, the
relational emergence of \emph{standard}, smooth manifold spacetime
is only guaranteed when $A$ is commutative. This opens the possibility
of having a new physical theory, with non-commutative $A$, in which
minimal Things and their properties are fundamental and assumed, but
where standard spacetime is \emph{forcefully} absent. That is, in
the usual approach, the ad-hoc assumptions given by the standard notions
of space and time give the usual algebraic structure to $C^{\infty}(M)$,
but, if we relax the former, we lose the manifold structure and therefore
the algebraic structure on the latter, while in the second approach,
one can start from an algebraic structure on the set of properties,
which will give rise to the manifold and the standard spacetime according
to the commutativty of this algebra (of course, in the theory of previous
sections, standard spacetime is also secondary and not fundamental,
since it's derived from minimal Things and their properties, which
are assumed to be the fundamentals of the ontology; nevertheless,
the axioms that make standard spacetime to appear relationally from
matter are ad-hoc, they can be put there but also removed without
any real justification more than that their assumption leads to the
desired result, while, in the approach discussed here, the emergence
of standard spacetime from minimal Things is forcefully forbidden
if $A$ is non-commutative\footnote{We are talking about a fully non-commutative case, not the ``almost
commutative'' one on which the spectral standard model is based,
composed of a tensor product of algebras, one which is the standard
commutative algebra of a manifold.}.)

$\;$

\emph{In this way, the main learning from all of this is that standard
spacetime is not only something ontologically non-fundamental because
it can be relationally derived from Things (as in the theory in the
previous section), but because there may be cases in which, even when
there are indeed Things, standard spacetime cannot even be relationally
derived from them, and, of course, much less substantivally so.}

$\;$

That $A$ may not be commutative is suggested by the quantization
of gravity. Indeed, its elements are the properties of (minimal) fields
and, when all fields including gravity are quantized (and, therefore,
one cannot use a classical Gravitational Field as background to build
a classical, commutative spacetime), these properties form non-commutative
algebras. Although, the relation between $A$ and the quantum field
algebras doesn't seem straightforward, since, in the classical case,
$M$ (and, therefore, $C^{\infty}(M)$) is built from a single solution,
while the algebras of functions on phase space (which are the ones
actually related to the quantum field algebras) have as domains the
phase space points comming from all solutions. Nevertheless, recall
that our relationist approach here\emph{\large{} }\emph{gives a way
of obtaining the algebra of physical $3-$space purely from Field
properties} (represented here by the phase space functionals), and,
therefore, it can be used as a guide to obtaining the corresponding
quantum counterparts, as done in \cite{key-4,key-5}. In any case,
this still shows that it seems unlikely that the quantum formalism
based on the quantum algebra of gravity could have some standard spacetime
as background (even the very standard manifold $M$, and not only
the metric, which is an obvious thing.)

$\;$

\subsubsection*{{\large{}Appendix: Elementary Philosophy of Physics (Ontology and
Semantics)} \cite{key-1}}

$\;$

\subsubsection*{A.1. Ontology }

$\;$

Physical theories assume that Reality \emph{is made of something}.
In ontology, this something is called a \emph{Substance}. In science,
it's assumed that Reality is made of a single type of Substance, which
is called \emph{Material Substance}. Therefore, science's ontology
is \emph{monistic}\footnote{Other metaphysical theories may introduce other types of Substances,
like ``Spatio-Temporal Substance'', ``Immaterial Souls'', ``Mental
Substance'' (this one is famously associated with Descartes and his
dualist, in terms of Substance, solution to the, also famous, Mind-Body
problem, i.e., what's the mind and how does it interact with the physical
body?), etc. But, in science, \emph{only matter interacts and associates
with matter}.}. In ordinary language, we refer to this Substance simply as \emph{Matter}.
Nevertheless, the ``matter'' of which we are talking about is not
necessarily some chunk of something with some mass, as, say, a piece
of steel. The notion of Matter used in science can become quite subtle
and unintuitive. In fact, the best and more up-to-date physical theories
should, and must, be the ones that \emph{dictate} the\emph{ specifics}
of the ontology and the particular properties of matter. A key property
of Material Substance is that two \emph{material individuals} can
\emph{Associate} to form a third material individual. If $a$ and
$b$ are these two material individuals, we will denote the association
operation as $\dotplus$, and, in this sense, if $c$ is the third
material individual mentioned before, we write (we will denote as
$S$ the set which contains our material individuals):

\[
c=a\dotplus b.
\]

That is, in science we have different types of ``matter'', which
are each characterized by the binary association property, $\dotplus:S\times S\,\longrightarrow S$,
among its elements\footnote{Mathematically, one can take a Boolean $\sigma-$algebra structure
for the triple $(S,\dotplus,\boxempty$) (where $\boxempty$ represents
the null element, or minimal element, $R$ the maximal element, see
below, and $\dotplus$ the ``disjunction'' or $sup$, that is, the
minimal upper bound between elements; one also has the $inf$ or ``conjunction'',
which is interpreted as the Interposition $\dot{\times}$ of material
individuals.)}. We postulate that all Material Substance carries \emph{Energy},
where this concept is defined according to the best physical theory
at disposition. In general, it's a Property (see below) of matter
which is represented, in one way or another, by an additive, real
function on matter (which depends on the reference frame adopted,
though), i.e., an $h(a)$ such that

\[
h(a\dotplus b)=h(a)+h(b),
\]

for any two elements $a,b$ from $S$, and which ``generates time
translations'' in the context of that theory (this is usually done
in what is called the ``Hamiltonian formulation'' of a dynamical
theory.) What this means is that \emph{Energy is the Property of a
Thing that allows us to determine, in a physical theory, how the Properties
of this Thing change in time} (see below for these terms.)

$\;$

Thus, if something doesn't posses energy, then it cannot be considered
matter.

$\;$

Of course, the things that inhabit Reality are much more than mere
\emph{undifferentiated} bunches of material substance. The ``fauna''
of Reality is very rich, with individuals that can have very peculiar
and different \emph{Substantial Properties}. A Substantial Property
is a property/quality that a material element \emph{possess} or \emph{bears}
and it's as objective as the existence of the material element itself
(properties are not substantial elements themselves, though) We will
call \emph{Thing} to a matter element which possess substantial properties.
Things also associate to produce other Things. We denote the set of
\emph{all} Things that exist as $\Theta$. We mention that a very
particular type of properties are the so-called \emph{Emergent Properties},
that is, properties which are present in the \emph{aggregate} Thing
$c=a\dotplus b$, but \emph{not} in its component Things, $a$ and
$b$, before the aggregation.

$\;$

Reality is the Thing $R\in\Theta$ which consists in the association
or aggregation of \emph{all} Things in $\Theta$.\footnote{One could say, what about time and space, are not they part of Reality
too? This is a complex issue. They are intrinsically tied to Things
in a certain specific sense, as we will see. Also, what about logic?
We use logic to understand Reality, but we don't take it as part of
Reality (it's not even a thing.)}

\subsubsection*{A.2. Semantics}

$\;$

\selectlanguage{spanish}%
A mathematical theory without a factual interpretation is only mathematical,
not physical. A scalar field satisfying the Laplace equation may be
the potential for a gravitational field or for a static electric field.
It is the factual interpretation (which interprets this field as the
potential for a gravitational field or for a static electric field)
the thing that transforms it into a theory of physics. The physical
interpretation maps mathematical objects\footnote{These mathematical objects are mere mental \emph{constructs} made
by human brains. In this way, we do not adhere to mathematical Platonism
(which says they exist in their ``own objective reality'') However,
this does not imply a free subjectivism, since different humans can
manage to understand each other through the same concepts. Then, in
practice, we can pretend as if Platonism were true.} to objects of the real world. The purely semantic map, $\varphi$,
assumes a certain kind of \emph{realism} (which is an ontological
hypothesis), simply the same that is needed to give meaning to the
fundamental maxim of the scientific method (we will consider only
classical realism here). In this way, the interpretation here is very
simple: factual item $y$ of this physical system will be represented
mathematically by, say, the mathematical function $f$ (which is a
construct.) We don't need anything else, we simply refer to the factual
item because it exists, it is simply there, we simply point to it
(semantically speaking, not pointing in a literal sense.) When a particular
mathematical model from a physical theory is interpreted in terms
of some physical notions, one says that these (often vague in the
quantitative sense) notions have been \emph{exactified}.

\selectlanguage{english}%
$\;$

\selectlanguage{spanish}%
The purely semantic interpretation takes a mathematical construct
and relates it to the real world. But this is not enough to give physical
meaning to this construct. The purely semantic interpretation gives
us, at best, the \emph{physical referents} of the construct, that
is, those physical \emph{entities} (i.e., the Things that comprise
the Reality hypothesized by the physical theory in consideration)
to which the physical interpretation refers. What is important to
note is that, in a full theory of physics, this construct is not isolated.
Indeed, there are also other mathematical constructs in the theory,
and, in general, they are all \emph{inter-related} with one another
(\emph{in the logical sense}.) These interrelations constitute what
is called the \emph{mathematical Sense} of the construct. Some of
these other constructs are also physically interpreted in terms of
purely semantic interpretations, and, thus, the mathematical sense
becomes also a \emph{physical sense}. In this way, \emph{given a physical
theory, we will adopt the point of view in which the meaning of a
construct is fully established by its total Sense and its Reference
class} ($\varSigma\ni\sigma$ and $\varSigma\subseteq\Theta_{Universe}$)
of Things ($\sigma$), \emph{both which can only be read once the
theory has been fully established in its mathematical axioms and semantic
interpretations}. Note that, in a completely axiomatized theory, certain
basic constructs will determine the meaning of the other constructs
of the theory (in general, these basic constructs will be those whose
physical interpretations are made in terms of factual elements that
are usually taken as factual primitives, such as the notions of length,
lapses of time, facts, propensity, etc.) Space and time, in particular,
are quite generic to most theories. In Scientific Ontology, one can
actually make theories that give them a \emph{precise} physical meaning.
These theories are appropriately called ``\emph{proto-physics}''
(to construct these theories for space and time will be the main aim
here) Although, one often needs to insert them into a physical theory
in order to have a more \emph{exact} (in a \emph{quantitative} sense)
meaning. 

$\;$
\selectlanguage{english}%

\end{document}